\documentclass{Interspeech2024}
\usepackage{amsmath,graphicx}
\usepackage{amssymb}
\usepackage{siunitx}
\usepackage{tikz}
\usetikzlibrary{arrows}
\usepackage{hyperref}
\usepackage{makecell}
\usepackage{booktabs}
\usepackage{multirow}

\interspeechcameraready

\title{tinyCLAP: Distilling Contrastive Language-Audio Pretrained models}

\name[affiliation={}]{Francesco}{Paissan}
\name[affiliation={}]{Elisabetta}{Farella}

\address{
  Fondazione Bruno Kessler, Italy}
\email{\{fpaissan, efarella\}@fbk.eu}

\keywords{contrastive language-audio pretraining, tinyML, sound event detection, zero-shot classification, distillation and pruning}

\begin{document}

\maketitle

\begin{abstract}
    Contrastive Language-Audio Pretraining (CLAP) became of crucial importance in the field of audio and speech processing. Its employment ranges from sound event detection to text-to-audio generation. However, one of the main limitations is the considerable amount of data required in the training process and the overall computational complexity during inference. This paper investigates how we can reduce the complexity of contrastive language-audio pre-trained models, yielding an efficient model that we call tinyCLAP. We derive an unimodal distillation loss from first principles and explore how the dimensionality of the shared, multimodal latent space can be reduced via pruning. tinyCLAP uses only 6\% of the original Microsoft CLAP parameters with a minimal reduction (less than 5\%) in zero-shot classification performance across the three sound event detection datasets on which it was tested.
\end{abstract}

\section{Introduction}

Contrastive Language-Audio Pretraining (CLAP) \cite{Elizalde2022CLAPLA}, and similarly, its image counterpart, CLIP \cite{Radford2021LearningTV}, proved to be an effective technique to pretrain audio and image encoders. In particular, CLAP (Figure \ref{fig:diagram}), and some of its variants \cite{Wang2023UnsupervisedIO, Wu2022LargescaleCL} achieved state-of-the-art performance for event detection, showcasing impressive results also in Zero-Shot (ZS) classification. To accomplish this, CLAP learns a similarity score between audio samples and text during the pretraining stage, which is then used to compute the affinity with unseen text prompts, potentially representing new classes. Due to its ability to correlate the audio and text modalities, CLAP finds applications in text-conditioned generative models as well, where a correlation between the text embedding and audio embedding is needed \cite{Ramesh2021ZeroShotTG, Liu2023AudioLDMTG}.

Reducing the computational complexity of CLAP poses significant challenges. Still, it can yield many benefits in acoustic scene classification and audio generation in resource-constrained devices, such as those in IoT scenarios. The model capacity needed for learning the correlations between audio and text is high. Thus, CLAP encoders are not suited for fast and low-footprint inference. To overcome this challenge, we explore the use of two standard network compression techniques, namely knowledge distillation \cite{Hinton2015DistillingTK} and pruning \cite{Blalock2020WhatIT}, as they proved to be effective methods for learning more compact models while inheriting the representation capabilities of the teacher model. Still, standard knowledge distillation~\cite{Hinton2015DistillingTK} is unsuitable for the CLAP audio encoder because its representations can be used in diverse scenarios, and the number of logits is not defined a priori. It means that the dimensionality of the soft labels can change depending on the downstream task, hindering the possibility of running the vanilla knowledge distillation scheme. Additionally, text data might not be available when adapting the CLAP weights for a specific application domain. Thus, we focus on developing distillation and pruning strategies that can work with audio samples only, hindering the need to use the corresponding caption.  

Previously, other approaches based on affinity mimicking and weight inheritance~\cite{wu2023tinyclip} tried to solve the problem of distilling models trained with a contrastive objective. However, these models (i) use cross-modal distillation schemes and (ii) produce networks that are too demanding for edge applications, counting up to tens of millions of parameters. To our knowledge, this manuscript is the first to (i) tackle the challenges of unimodal distillation of the CLAP audio encoder and pruning of the shared multimodal latent space, together with (ii) presenting low-footprint networks for zero-shot classification. We want to emphasize that we used the original CLAP weights\footnote{\url{https://zenodo.org/records/8378278}} during the experimental evaluation of this manuscript for convenience. However, this approach can be used with all the other CLAP and CLIP variants and similarly formulated ZS classification models.

Section \ref{sec:methods} presents the methods proposed in the paper, providing the elements needed to understand CLAP and our contribution. Then, in Section~\ref{sec:exp}, we describe the experimental setup used for the evaluation of the proposed methods. Finally, we present the results in Section \ref{sec:results}. We release the code for our submission through a companion website\footnote{\url{https://fpaissan.github.io/tinyclapweb/}}.


\section{Methods}
\label{sec:methods}
\begin{figure*}[t!]
    \hspace{-0.4cm}
    \centering
    \includegraphics[width=0.82\textwidth]{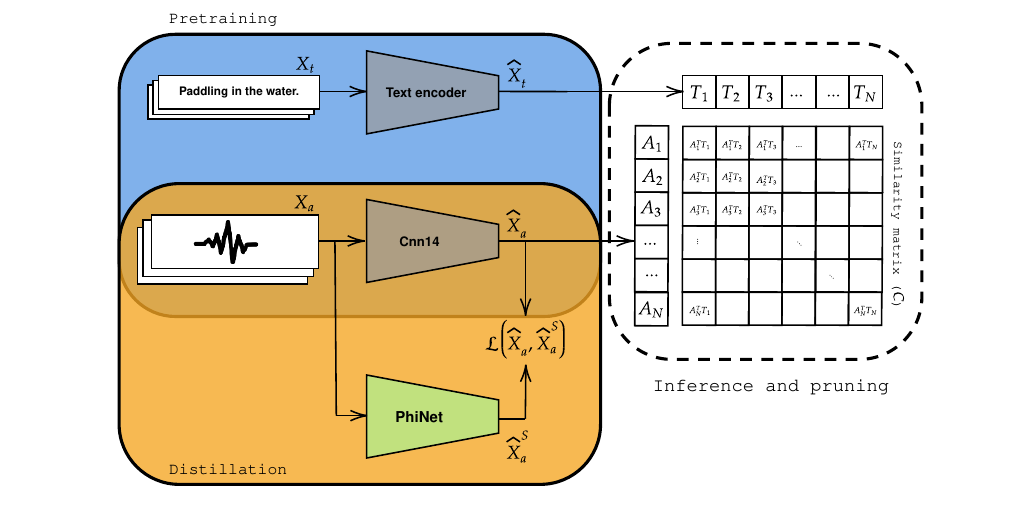}
    \caption{Diagram of the proposed distillation technique. During distillation, the student and audio encoders are aligned by minimizing the loss in Eq. \ref{eq:loss}. During the distillation stage, the encoders represented in grey are frozen, while those in green are trained. For simplicity, the image does not include the projection layers.}
    \label{fig:diagram}
\end{figure*}

This section introduces CLAP's core principles and the notation used throughout the paper. It also presents the distillation process and elaborates on how the pruning is performed. Figure \ref{fig:diagram} shows an overview of the proposed method.

\subsection{Contrastive Language-Audio Pretraining}
\label{sec:clap}
Let $\mathbf{X}_a$ be the pre-processed input audio such that $\mathbf{X}_a \in \mathbb{R}^{F\times T}$, where $F$ is the number of spectral components (e.g. mel bins), and $T$ is the number of time frames. Let $\mathbf{X}_t$ represent the corresponding text caption. Given a set of $N$ elements, $\{\mathbf{X}_a, \mathbf{X}_t\}_{i=1}^N$ represents a batch of audio-text pairs. For convenience, we will refer to the batch as $\{\mathbf{X}_a, \mathbf{X}_t\}$ throughout the manuscript. Text and audio features are extracted via text and audio encoders. Let $f_a(\cdot)$ be the audio encoder and $f_t(\cdot)$ be the text encoder, then for a batch:
\begin{equation}
    \hat{\mathbf{X}}_a = f_a(\mathbf{X}_a)\text{; } ~~ \hat{\mathbf{X}}_t = f_t(\mathbf{X}_t),
\end{equation}

where $\hat{\mathbf{X}}_a \in \mathbb{R}^{N\times U}$ and $\hat{\mathbf{X}}_t \in \mathbb{R}^{N\times V}$ with $U$, $V$ the audio and text latent dimensions respectively.

After encoding, CLAP brings the audio and text representations to a joint multimodal space of dimensionality $d$ using a learned linear projection:
\begin{equation}
    \mathbf E_a = \mathbf L_a(\hat{\mathbf X}_a)\text{; } ~~ \mathbf E_t = L_t(\hat{\mathbf X}_t),
\end{equation}
where $\mathbf E_a \in \mathbb{R}^{N\times d}$, $\mathbf E_t \in \mathbb{R}^{N\times d}$, and $L_i$ represent the linear projection layers.

In this shared latent space, we can measure the similarity between audio and text pairs:
\begin{equation}
    \mathbf C = \tau \mathbf E_t \mathbf E_a^\top.
\end{equation}

Note that this is a scaled cosine similarity between each other element in the original batch of $N$ elements, with temperature $\tau$. The CLAP loss optimizes the audio and text encoders and the linear projection layers to maximize the normalized similarity of positive pairs on the diagonal of matrix $\mathbf C$, while minimizing the similarity of the negative pairs.

\subsection{Distilling Audio Representations without Text}
\label{sec:distill}
Let $f_a^S(\cdot)$ be the distilled audio encoder, and $L_a^S(\cdot)$ be the linear projection from the student embedding to the shared multimodal latent space, then:
\begin{equation}
    \hat{\mathbf X}_a^S = f_a^S(\mathbf X_a); ~~ \mathbf E_a^S = L_a^S(\hat{\mathbf X}_a^S)
\end{equation}
where $\hat{\mathbf X}_a^S \in \mathbb{R}^{N\times U}$ are the audio features extracted from the student network and $\mathbf E_a^S \in \mathbb{R}^{N\times d}$ is the projection in the shared latent space.

To distill the audio encoder $f_a(\cdot)$ to a more computationally efficient network ($f_a^S$), one option is to follow the standard knowledge distillation approach \cite{Gou2020KnowledgeDA}. However, this method would require using both the audio and text modality to compute the soft labels for a fixed number of classes, thus limiting the generalization capabilities of the models. Instead, focusing on the structure of the latent space, we can distill the audio encoder using a single modality (e.g. only audio or text) as the projections are in a shared latent space. We recall that the cosine similarity is maximum at convergence for positive audio-text pairs. This means that the two projections are aligned and in the same direction.

Then for a perfectly distilled model, we have:
\begin{equation}
    \label{eq:distill0}
    \cos(\mathbf E_a, \mathbf E_t) = 1 \Leftrightarrow \cos(\mathbf E_a^S, \mathbf E_t) = 1.
\end{equation}

Geometrically, this means that $\mathbf E_a$, $\mathbf E_t$, and $\mathbf E_a^S$ are aligned and in the same direction. Thus, also
\begin{equation}
    \label{eq:distill}
    \cos(\mathbf E_a, \mathbf E_a^S) = 1
\end{equation}
holds.
We want to note that the implication between Eq. \ref{eq:distill0} and Eq. \ref{eq:distill} does not have an analytical proof as the cosine distance does not satisfy the triangle inequality \cite{Schubert2021ATI}. However, our empirical evaluation shows that this approximation holds for the scope of this paper (see Sec. \ref{sec:sanity}).

Eq. \ref{eq:distill} directly defines a cost function that maximizes the cosine similarity between the new and original audio encoders, which can be used for the distillation process:
\begin{equation}
    \label{eq:loss}
    \mathcal{L} = -\mathbf E_a^S~\text{sg}[\mathbf E_a^\top],
\end{equation}
where $\text{sg}[\cdot]$ represents the stop-gradient operation. As mentioned, this loss function requires only the audio modality for the distillation process.

\begin{table*}[ht!]
\centering
\newcolumntype{C}{>{\centering\arraybackslash}p{0.09\textwidth}}
\newcommand{\SplitCenter}[1]{%
  \begingroup
  \renewcommand{\arraystretch}{1.0}%
  \setlength{\tabcolsep}{0pt}%
  \begin{tabular}[t]{@{}>{\centering\arraybackslash}p{0.1\textwidth}@{}>{\centering\arraybackslash}p{0.1\textwidth}@{}}
    \hspace{0pt} & \hspace{0pt} \\
    \multicolumn{1}{c}{#1} & 
  \end{tabular}%
  \endgroup
}
\renewcommand{\arraystretch}{1.15}
    \caption{Tested networks for distillation. Hyperparameters $\alpha, \beta, t_0, N$ were derived for the different computational budgets using Hardware Aware Scaling. The last row shows the distillation results for the original CLAP encoder. Parameter count is reported for the models during the distillation process, therefore with $r=1024$. The last row presents the results for the original CNN14 CLAP encoder without distillation.}
\begin{tabular}{cccccSSSS}
    \toprule
                   &      &      &   &   &       & \multicolumn{3}{c}{\textbf{ZS Accuracy (\%)}}      \\ \cmidrule{7-9}
    \textbf{Student model} & \boldsymbol{$\alpha$} & \boldsymbol{$\beta$} & \boldsymbol{$t_0$} & \boldsymbol{$N$} & \textbf{Params {[}M{]}} & \text{TUT17} & \text{US8k} & \text{ESC50} \\ \midrule
    PhiNet\_1      & 3.00 & 0.75 & 6 & 7 & 7.0  & 25.2 & 68.3  & 77.4  \\
    PhiNet\_2      & 3.00 & 0.75 & 6 & 9 & 13.0 & 26.4 & 69.7  & 77.2  \\
    PhiNet\_3      & 3.00 & 0.75 & 4 & 7 & 6.2  & 26.1 & 70.3  & 76.5  \\
    PhiNet\_4      & 1.50 & 0.75 & 6 & 7 & 4.4  & 27.5 & 67.9  & 73.0  \\
    PhiNet\_5      & 0.75 & 0.75 & 4 & 7 & 3.5  & 26.7 & 65.2  & 66.1 \\
    PhiNet\_6      & 0.75 & 0.75 & 4 & 4 & 3.2  & 22.1 & 51.8 & 41.9 \\
    PhiNet\_7      & 0.75 & 0.75 & 6 & 4 & 3.3  & 22.3 & 51.6 & 44.1 \\ \midrule \midrule
    CNN14 & /    & /    & / & / & 82.8 & 28.9  & 72.1  & 82.3  \\
    CNN14-CLAP & /    & /    & / & / & 82.8 & 29.6 & 73.2 & 82.9 \\
    \bottomrule
    \end{tabular}
    \label{tab:results}
\end{table*}

\subsection{Pruning the Shared Multimodal Latent Space}
\label{sec:pruning}

To compute the loss function and distill the audio encoder using the shared multimodal latent space, it is essential that the two projections maintain the same dimensionality $d$. Therefore, the output of the projection layers should always be $d$, precluding parameter or operation reduction. 
After the distillation process, however, the dimensionality of the shared latent space can be reduced to $r < d$. The similarity metric measures the alignment between vectors. Therefore, zeroing out the entries associated with the smallest absolute value in the vector negligibly affects the vectors' direction and, consequently, the structure of the latent space. 

To avoid data leakage, we rank the vector entries on the training set. Therefore, we compute the average absolute value of the student projections outputs $\mathbf E_a^S$ on the entire training set, such that we can rank the latent dimensions from the most important to the least, following:

\begin{equation}
    I = \text{argsort}\left(\frac{1}{D}\sum_{i=1}^D | \mathbf E_{a,i}^S | \right)
\end{equation}
where $D$ is the number of samples in the training set, $\mathbf E_{a,i}^S$ is the projection of the $i$-th training sample in the shared, multimodal latent space, $|\cdot|$ is the element-wise absolute value operation, and the $\text{argsort}(\cdot)$ returns the indexes such that the entries of $\mathbf E_{a,i}^S[I]$ are sorted in a decreasing order.

\subsection{Efficient Audio Encoder}
To reduce the computational complexity of the original CLAP encoder, we used PhiNet \cite{Paissan2021PhiNetsAS}, a scalable backbone for edge processing. PhiNet is based on inverted residual blocks. It was originally developed for computer vision applications and then benchmarked on audio processing tasks \cite{Brutti2022OptimizingPA, Paissan2022ScalableNA, 10164328}, showcasing good performance-complexity trade-offs. In the same paper, the authors proposed Hardware-Aware Scaling (HAS), a network scaling procedure that maps the computational requirements of the networks to their hyperparameters. For the purpose of this manuscript, we used HAS as a scaling technique to benchmark zero-shot classifiers with different computational budget targets. 



\section{Experimental setup}
\label{sec:exp}


\textbf{Pre-processing.} Following the original CLAP implementation, we re-sampled all the audio files to \SI{44.1}{\kilo\hertz}. We computed Mel spectrograms with 64 Mel bins, a window size of 1024 samples and a hop size of 320 samples. Before feeding the samples through the neural network, we normalized the spectrograms along the frequency axis.\\
\textbf{Encoders.} As the teacher model, we used Microsoft CLAP. Its encoders are a CNN14 \cite{Kong2019PANNsLP} for audio encoding and a BERT transformer \cite{Devlin2019BERTPO} for text encoding. After validating the distillation loss on a self-distillation experiment, where the teacher and student network are the same CNN14, we used PhiNets \cite{Paissan2021PhiNetsAS} with different hyperparameter configurations for the student model. The choice of hyperparameters was guided by the Hardware-Aware Scaling (HAS) principle \cite{Paissan2021PhiNetsAS} to match a variety of computational budgets. This approach allows for a proper benchmarking of how the performance of the ZS classifiers changes with the computational complexity of the encoders. Table \ref{tab:results} summarizes the configurations of the students and their relative ZS classification performance. \\ 
\textbf{Distillation.} For the distillation process, we used the audio samples from the same datasets as the original Microsoft CLAP paper \cite{Elizalde2022CLAPLA}, namely AudioCaps \cite{Kim2019AudioCapsGC}, MACS \cite{Morat2021MACSM}, FSD50k \cite{Fonseca2020FSD50KAO}, and ClothoV2 \cite{Drossos2019ClothoAA}. Each waveform was randomly truncated during training to a continuous segment of \SI{5}{\s}. The captioning process was discarded, as the distillation loss only requires audio samples. For training, we used a two-stage approach. First, we used the Adam optimizer with the loss function defined in Eq.~\ref{eq:loss} and a learning rate of \num{3e-3} to update the weights of $f_a^S(\cdot)$ and $L_a^S(\cdot)$ simultaneously for \num{100} epochs. Afterwards, we used the same setup, but with a learning rate of \num{1e-3} to update the projection layer only ($L_a^S$). Note that the stop-gradient operator in Eq.~\ref{eq:loss} means we do not update the teacher weights during the distillation process.\\
\textbf{Zero-Shot Evaluation.} For evaluating the Zero-Shot (ZS) performance of the distilled model, we used three acoustic scene classification datasets, namely TUT17 \cite{Drossos2019LanguageMFTUT}, ESC50 \cite{piczak2015dataset} and US8k \cite{Salamon2014ADA}. Similarly to many ZS benchmarking setups, we prepended the sentence \textit{`this is the sound of '} to the labels in order to define the caption. The text representations and projections are computed using the frozen text encoder ($f_t$) and projection layers ($L_t$) of the CLAP teacher model. As described in Section \ref{sec:pruning}, at inference time, we use the dictionary $I$ to only keep the top-$r$ projection vector entries (referred to as $I^{(r)}$ from now on) of both the text and student audio encoders. As per CLAP,  the class probabilities are computed with:
\begin{equation}
    p = \text{softmax}(\mathbf E_a^S[I^{(r)}]\cdot \mathbf E_t^c[I^{(r)}]^\top)
\end{equation}
where $\mathbf E_t^c$ refers to the projection in the shared latent space of the text caption corresponding to class $c$.
\section{Results}
\label{sec:results}

\tikzstyle{dictsmall} = [draw, thick, fill=white!10, rectangle, 
    minimum height=1.0cm, minimum width=5cm] 
    \newcommand{\xshifts}{+4.7}
\begin{figure*}[t!]
    \centering
    \resizebox{17.cm}{!}{
    \begin{tikzpicture}[auto, node distance=1.2cm]
        \node [draw=none, fill=none] (latents_all)  { \includegraphics[scale=0.5]{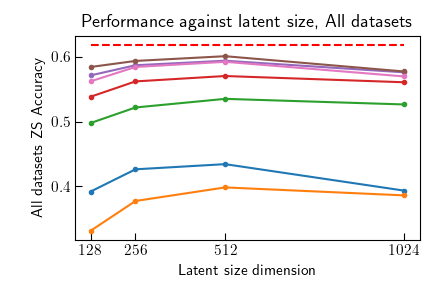} };
        \node [draw=none, fill=none, right of=latents_all,  xshift=8cm] (comp_complex)  { \includegraphics[scale=0.7]{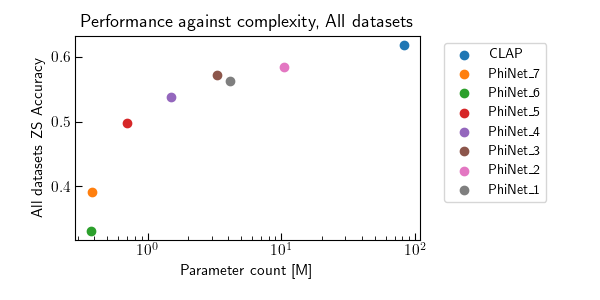} };
        
        \node [draw=none, left of=latents_all, xshift=-3.7cm, yshift=0.3] (legend) { \includegraphics[height=3.2cm]{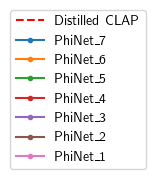}};
        \node [draw=none, fill=none, below of=latents_all, yshift=-1.5cm] (desc1)  { \small a) };
        \node [draw=none, fill=none, below of=comp_complex, yshift=-1.5cm] (desc2)  { \small b) };
    \end{tikzpicture}
    }
    \vspace{-0.25cm}
    \caption{a) Impact of the latent space pruning on the distilled checkpoints. The pruning procedure is described in Sec \ref{sec:pruning}. b) Computational complexity against ZS performance of the distilled models. Note that the performance of the distilled and the original models match, considering the baseline distillation scenario. Similar plots, computed for all datasets independently, can be found on the companion website.}
    \vspace{-0.5cm}
    \label{fig:results}
\end{figure*}

The experimental results are divided into two sections. The first focuses on validating the distillation process, whilst the second one showcases how the ZS performance of distilled models drops with varying computational budgets. Due to space limitations in the manuscript, the figures showcase the average of the ZS performance for each task independently, and the tables do not explicitly mention the parameter count for all latent size combinations. We invite the reader to refer to the companion website for these additional resources.\footnote{\url{https://fpaissan.github.io/tinyclapweb/}}

\vspace{-0.1cm}
\subsection{Distillation Sanity Checks}
\label{sec:sanity}
To validate the distillation and pruning strategies, we used the same audio encoder of the original CLAP implementation as the student encoder. This approach enables validating the distillation process itself by isolating possible contributions related to model capacity and regularization. The results of this analysis are reported in Table \ref{tab:results}. On the three benchmarks employed in this study, the distilled CLAP encoder (CNN-14 in Table \ref{tab:results}) achieved comparable results with the original model. In fact, the ZS performance dropped only \num{0.8}\%, averaged on the three benchmarks. Note that this slight fluctuation can also depend on the CUDA and PyTorch version or minor changes in the hyper-parameters. Since it was outside the scope of this manuscript and computationally intensive, we have not performed a hyperparameter search on the training parameters of the distillation process for this network. Finally, since the distilled model performed as well as the original one, we consider the distillation process successful and Eq. \ref{eq:loss} validated as a cost function.

\vspace{-0.1cm}
\subsection{Reducing Computational Complexity}
As described in Sec. \ref{sec:methods}, CLAP computational complexity can be reduced with a more efficient audio encoder and a latent space pruning strategy. Hereafter, we analyze the individual contributions of these two strategies to the performance-complexity trade-off.
\\
\textbf{Audio Encoders.} As shown in Table \ref{tab:results} and Figure \ref{fig:results}, the networks range between \num{13}M and \num{3.2}M parameters, without accounting for the latent space pruning, and its consequent reduction in model size. Overall, the performance of the models drops significantly with decreasing number of parameters. For PhiNet, as expected, even the slightest increase in parameter count correlates to an increase in ZS performance. This result aligns well with the choice of using HAS, which accommodates the biggest network for each computational budget. Quantitatively, for a reduction of approximately 4\% of ZS classification performance across all benchmarks, PhiNet uses only 8\% of the parameters of the original encoder without pruning the latent space. This result refers to the hyperparameters of the model denoted as PhiNet\_3 in Table~\ref{tab:results}. \\
\textbf{Latent Space Pruning.} The impact of the latent space dimensionality reduction on the overall complexity of the network is related to the structure of the projection layers. Specifically, the two linear projection layers can be streamlined to output only the pertinent entries for each model, either through manual adjustments or utilizing approaches such as \cite{Fang2023DepGraphTA}. 
The impact of this complexity reduction technique is highlighted in Fig. \ref{fig:results}a), where the average performance on the three benchmarks is showcased with respect to the size of the latent space. It is worth emphasizing that these models are not trained from scratch with a specific latent space dimensionality; rather, they are derived from the distilled checkpoints of Table \ref{tab:results} following the methodology described in Sect. \ref{sec:pruning}. A clear trend emerges, indicating that for a latent space dimensionality of $r=512$, the ZS classification performance is greater or equal to the case of $r=1024$. Therefore, the model's complexity is diminished by approximately \num{40}\% for the tested networks without compromising the classification performance. 


\section{Conclusions}

In conclusion, this paper presented a distillation technique for CLAP and a pruning strategy for its latent space that only uses the audio modality. We validated the results on three public acoustic scene classification benchmarks. Our tinyCLAP model demonstrates its versatility by successfully adapting to models of varying computational complexities.
We achieved considerable compression in model size (only 6\% of the original parameters), with a minor drop in ZS classification accuracy (around 4\%, averaged on all benchmarks). The proposed approach can be successfully applied to different CLAP variants and can extend to CLIP as well.

\section{Acknowledgements} We acknowledge the support of the PNRR project FAIR -  Future AI Research (PE00000013),  under the NRRP MUR program funded by the NextGenerationEU.

\newpage

\bibliographystyle{IEEEtran}
\bibliography{mybib}

\onecolumn
\appendix

\section{Results on Entire Evaluation Setup}

In Table~\ref{tab:full}, we report the results obtained using all combinations of models and pruning factor, $r$. In the official submission, these results are available through our companion website.

\begin{table}[h]
    \centering
    \caption{Detailed results.}
    \begin{tabular}{ccSccc}
    \toprule
     &  & \textbf{Params {[}M{]}} & \textbf{ESC-50} & \textbf{UrbanSound8K} & \textbf{TUT17} \\
     \midrule
    \multicolumn{1}{c|}{\multirow{7}{*}{\rotatebox{90}{$r = 1024$}}} & PhiNet\_7 & 3.3 & 44.1 & 51.6 & \multicolumn{1}{c}{22.3} \\
    \multicolumn{1}{c|}{} & PhiNet\_6 & 3.2 & 41.9 & 51.8 & \multicolumn{1}{c}{22.1} \\
    \multicolumn{1}{c|}{} & PhiNet\_5 & 3.5 & 66.1 & 65.2 & \multicolumn{1}{c}{26.7} \\
    \multicolumn{1}{c|}{} & PhiNet\_4 & 4.4 & 73.0 & 67.8 & \multicolumn{1}{c}{27.5} \\
    \multicolumn{1}{c|}{} & PhiNet\_3 & 6.2 & 76.5 & 70.3 & \multicolumn{1}{c}{26.1} \\
    \multicolumn{1}{c|}{} & PhiNet\_2 & 13.0 & 77.2 & 69.7 & \multicolumn{1}{c}{26.4} \\
    \multicolumn{1}{c|}{} & PhiNet\_1 & 7.0 & 77.5 & 68.3 & \multicolumn{1}{c}{25.2} \\
    \midrule
    \multicolumn{1}{c|}{\multirow{7}{*}{\rotatebox{90}{$r = 512$}}} & PhiNet\_7 & 1.4 & 49.7 & 54.9 & \multicolumn{1}{c}{25.7} \\
    \multicolumn{1}{c|}{} & PhiNet\_6 & 1.4 & 43.4 & 53.9 & \multicolumn{1}{c}{22.2} \\
    \multicolumn{1}{c|}{} & PhiNet\_5 & 1.7 & 67.2 & 66.8 & \multicolumn{1}{c}{26.6} \\
    \multicolumn{1}{c|}{} & PhiNet\_4 & 2.6 & 72.5 & 68.2 & \multicolumn{1}{c}{30.5} \\
    \multicolumn{1}{c|}{} & PhiNet\_3 & 4.3 & 77.4 & 71.1 & \multicolumn{1}{c}{29.8} \\
    \multicolumn{1}{c|}{} & PhiNet\_2 & 11.5 & 78.0 & 71.8 & \multicolumn{1}{c}{30.6} \\
    \multicolumn{1}{c|}{} & PhiNet\_1 & 5.2 & 77.9 & 69.2 & \multicolumn{1}{c}{30.7} \\
    \midrule
    \multicolumn{1}{c|}{\multirow{7}{*}{\rotatebox{90}{$r = 256$}}} & PhiNet\_7 & 0.7 & 47.8 & 53.8 & \multicolumn{1}{c}{26.3} \\
    \multicolumn{1}{c|}{} & PhiNet\_6 & 0.7 & 40.7 & 50.0 & \multicolumn{1}{c}{22.5} \\
    \multicolumn{1}{c|}{} & PhiNet\_5 & 1.0 & 65.9 & 65.8 & \multicolumn{1}{c}{24.9} \\
    \multicolumn{1}{c|}{} & PhiNet\_4 & 1.9 & 71.2 & 67.0 & \multicolumn{1}{c}{30.5} \\
    \multicolumn{1}{c|}{} & PhiNet\_3 & 3.5 & 76.8 & 70.1 & \multicolumn{1}{c}{29.3} \\
    \multicolumn{1}{c|}{} & PhiNet\_2 & 10.7 & 77.0 & 70.7 & \multicolumn{1}{c}{30.5} \\
    \multicolumn{1}{c|}{} & PhiNet\_1 & 4.5 & 77.0 & 68.1 & \multicolumn{1}{c}{30.3} \\
    \midrule
    \multicolumn{1}{c|}{\multirow{7}{*}{\rotatebox{90}{$r = 128$}}} & PhiNet\_7 & 0.4 & 46.1 & 50.5 & \multicolumn{1}{c}{21.0} \\
    \multicolumn{1}{c|}{} & PhiNet\_6 & 0.4 & 36.3 & 45.2 & \multicolumn{1}{c}{17.9} \\
    \multicolumn{1}{c|}{} & PhiNet\_5 & 0.7 & 64.9 & 62.2 & \multicolumn{1}{c}{22.3} \\
    \multicolumn{1}{c|}{} & PhiNet\_4 & 1.5 & 69.5 & 62.2 & \multicolumn{1}{c}{22.3} \\
    \multicolumn{1}{c|}{} & PhiNet\_3 & 3.3 & 75.6 & 68.1 & \multicolumn{1}{c}{27.8} \\
    \multicolumn{1}{c|}{} & PhiNet\_2 & 10.5 & 75.9 & 68.9 & \multicolumn{1}{c}{30.6} \\
    \multicolumn{1}{c|}{} & PhiNet\_1 & 4.1 & 74.6 & 65.5 & \multicolumn{1}{c}{28.6} \\
    \bottomrule
    \end{tabular}
    \label{tab:full}
\end{table}

\end{document}